\documentclass[12pt]{article}
 \usepackage{graphicx}
 \date{}
 \newcommand{\be}{\begin{equation}}
 \newcommand{\ee}{\end{equation}}
 \newcommand{\bear}{\be\begin{array}}
 \newcommand{\bea}{\begin{eqnarray}}
 \newcommand{\eea}{\end{eqnarray}}
 \newcommand{\dst}{\displaystyle}
 \newcommand{\fr}[2]{\frac{{\dst #1}}{{\dst #2}}}
 \newcommand{\ggam}{\mbox{$\gamma\gamma\,$}}
 \newcommand{\bm}{\boldmath}

 \newcommand{\epe}{\mbox{$e^+e^-\,$}}
 \newcommand{\egam}{\mbox{$\gamma e\,$}}

 \newcommand{\SM}{${\cal S} {\cal M}\;$}
 \newcommand{\DSM}{$2{\cal H} {\cal D}  {\cal M}$}
 \newcommand{\MSSM}{${\cal M}{\cal S} {\cal S}  {\cal M}$}
 \newcommand{\CP}{${\cal C} {\cal P}\;$}
 \newcommand{\fn}[1]{\footnote{ #1}}
 
 \newcommand{\egeh}{\mbox{$e\gamma\to e H\,$}}

 \def\lsi{\raise0.3ex\hbox{$<$\kern-0.75em\raise-1.1ex\hbox{$\sim$}}}
 \def\gsi{\raise0.3ex\hbox{$>$\kern-0.75em\raise-1.1ex\hbox{$\sim$}}}
 \newcommand{\lsim}{\mathop{\lsi}}

 \newenvironment{Itemize}{\begin{list}{$\bullet$}%
 {\setlength{\topsep}{0.2mm}\setlength{\partopsep}{0.2mm}%
 \setlength{\itemsep}{0.2mm}\setlength{\parsep}{0.2mm}}}%
 {\end{list}}
 \newcounter{enumct}

 \title {\bf \bm\CP ODD ANOMALOUS INTERACTIONS OF HIGGS BOSON IN
 ITS
 PRODUCTION AT PHOTON COLLIDERS}

 \author{I.F. Ginzburg$^{1,2}$\thanks{E-mail:
 ginzburg@math.nsc.ru},
 I.P. Ivanov$^{1,2,3}$\thanks{E-mail: i.ivanov@fz-juelich.de}\\
 \makebox[8cm][l]{\normalsize
 $^1$  Institute of Mathematics, Novosibirsk, Russia}\\
 \makebox[8cm][l]{\normalsize
 $^2$  Novosibirsk State University Novosibirsk, Russia}\\
 \makebox[8cm][l]{\normalsize
 $^3$  IKP, Forschungszentrum J\"ulich, Germany}\\
 }

\begin{document}

 \maketitle
 \vspace{-9cm}
 \makebox[\textwidth][r]{\large\bf FZJ-IKP(Th)-2000/07}
 \vspace{7.5cm}

 \begin{abstract}
We discuss the potentialities of the study of \CP odd
interactions of the Higgs boson with photons via its production
at $\ggam$ and $e\gamma$ colliders. Our treatment of $H\ggam$
and $HZ\gamma$ anomalous interactions includes a set of free
parameters, whose impact on physical observables has not been
considered before.  We focus on two reactions --- $\ggam \to H$
and \egeh --- and introduce the polarization and/or azimuthal
asymmetries that are particularly sensitive to specific features
of anomalies.  We discuss the ways of disentangling effects of
physically different parameters of anomalies and estimate what
magnitude of \CP violating phenomena can be seen in these
experiments.
\end{abstract}

 \section {Introduction}

In paper \cite{anom1} we studied potentialities to discover \CP
even anomalous interactions of the Higgs boson via its
production at \ggam\ and \egam colliders. Below we bring under
analysis effects of \CP-parity violating anomalies. They result
in the polarization and azimuthal asymmetries in the Higgs boson
production. With new opportunities for variation of photon
polarization at Photon Colliders \cite{kotser}, the Higgs
boson production at \ggam\ and \egam\ colliders has an
exceptional potential in the extraction of these anomalies. To
some extent, some similar issues have been considered in
Refs.~\cite{GG}--\cite{Asak}.  However, in the analysis there
the polarization potential was not used in its complete form and
some natural degrees of freedom in the parameter space were not
considered.  Besides, the authors cited consider either \ggam\
or \egam\ collisions separately.  In the present paper we have
in mind that experiments in \ggam\ and \egam\ collider modes
will supplement each other and provide complementary
opportunities in investigating Higgs boson anomalous
interactions.

In our analysis we assume the Higgs boson to be
discovered by the time the photon collider starts operating, so that
its basic properties will be known by that time. For the
definiteness, we assume that the Higgs boson coupling constants
will be found experimentally to lie close to their SM values.
Our substantial idea is the necessity of step-by-step strategy in
studying anomalous effects.  Namely, the first step is the study
of $H\ggam$ anomalies in \ggam collisions and the second step is
using \egam\ collisions for the study of $HZ\gamma$ anomalies
assuming $H\ggam$ anomalies (both \CP even and \CP odd) to be
studied at the first stage (with higher accuracy than it is
possible at the second stage).

The \ggam\ and \egam\ colliders will be the specific modes of
the future Linear Colliders (in addition to the $e^+e^-$ mode)
with the following typical parameters \cite{GKST,SLACDESY} ($E$
and ${\cal L}_{ee}$ are the electron energy and luminosity of
the basic \epe\ collider).

\begin{Itemize}
 \item {\em Characteristic photon energy $E_\gamma\approx 0.8E$.}
 {\em
 \item Annual luminosity is
 typically ${\cal L}_{\ggam}\approx 200$ fb$^{-1}$.
\item Mean energy spread $\langle\Delta E_\gamma\rangle \approx
0.07E_\gamma$.
\item Mean photon helicity $\langle\lambda_\gamma\rangle \approx
0.95$ with variable sign \cite{GKST}.
 \item Circular polarization of photons can be transformed into the
 linear one \cite{GKST,kotser}.}
 \end{Itemize}
The effective photon spectra for these colliders are given in
Ref.~\cite{GKot}. With the above properties, considering photon
beams at the Photon Collider as roughly monochromatic is good
approximation for our purposes.

The value of effects which can be observed in experiment is
given by the expected accuracy in the measuring of the cross
sections under interest.  For the \ggam colliders the expected
accuracy in the measuring of Higgs boson decay width will be 2\%
or better \cite{JikS}. For \egeh process we assume the
achievable accuracy to be $5\div 10$\%.

Throughout the paper we denote by $\lambda$ and $\zeta/2$ the
average helicities of photons and electrons and by $\ell$ the
average degree of the photon linear polarization. We use some \SM
notations: $s_W=\sin\theta_W$, $c_W=\cos\theta_W$, $v_e= 1-
4\sin^2\theta_W$ and $v=246$ GeV (Higgs field v.e.v.). In the
numerical calculations we assume the Higgs boson to lie in the
most expected mass interval 110--250 GeV.  Some further notation
is borrowed from ref. \cite{anom1}.

\section{Sources of \CP violation. Parameterization}

We consider below triple Higgs boson anomalous interactions
$H\gamma\gamma$ and $HZ\gamma$ in the processes $\ggam\to H$
and $\egam\to eH$. The quartic interactions lie beyond our
scope.

One can imagine two possible mechanisms of \CP violation in the
interactions of the Higgs boson. First, the observed Higgs boson
can be a mixture of purely scalar and pseudoscalar fields, as it
can happen in the multi-doublet Higgs models or in \MSSM, see
for details, e.g. \cite{GG,BGK} and Sec.~\ref{secmix} as an
example. In this case \CP violating effects could be either weak
or strong.

The second possibility is that the Higgs boson itself is \CP
even fundamentally but underlying interactions can break the \CP
parity conservation law. In this case we expect small \CP
violating effects in the interactions of Higgs boson with known
particles. In turn, this type of \CP violation can be caused
either by effects in the underlying theory, similar to the
aforementioned mixing, or by fundamental effects related, for
example, to the breaking of unitarity of $S$-matrix at very small
distances. (In the latter case the \CP breaking can originate,
in principle, from the possibility that the $S$-matrix is
unitary only when written it in terms of {\em observable
asymptotic states} and the unitarity appears broken if the space
of states is expanded by adding the {\em unobservable} unstable
$H$ final states.)

A natural question then arises, namely, whether we can
distinguish between these two possible causes of \CP violation:
i.e. {\em whether the energy scale of \CP violation $\Lambda$ is
low or high}.  In order to answer this question, one should
study how corresponding amplitudes depend on additional
kinematical variables, such as total energy $\sqrt{s}$, photon
virtualities $Q^2$ etc., i.e. on $Q^2/\Lambda^2$, $s/\Lambda^2$,
etc. Indeed, in the first case the dependence on these
parameters could be observable, while in the second case the
above dimensionless parameters are small and the corresponding
amplitudes appear independent from these kinematical variables.
(The latter case is described usually with the aid of effective
lagrangians.) However, a specific feature of reaction $\ggam\to
H$ is that its kinematics is fixed. This makes it impossible to
observe any additional dependence on $\Lambda$. As we turn to
process \egeh, one kinematical degree of freedom appears,
namely, the virtuality $Q^2$ of the exchanged photon or $Z$.
However, as shown in \cite{anom1}, the bulk of the cross section
comes from region $Q^2/M_H^2 \ll 1$, which again leaves us
unable to learn about the source of \CP violation.

The outcome of this discussion can be summarized as follows:
{\em when considering real Higgs boson production in
the two processes discussed, the above two sources of \CP violation
are indistinguishable in the discussed experiments}.

Given this, we follow a natural procedure to describe the
deviation of discussed production amplitudes from their \SM values
in a universal manner. We parameterize the $H\ggam$ and $HZ\gamma$
amplitudes (which will be also referred to as effective $H\ggam$
and $HZ\gamma$ vertices) in the operator form, similar to that
for the effective lagrangian:
\bear{c}
{\cal M}_{\gamma\gamma H} = \fr{1}{v}\left[G_\gamma
HF^{\mu\nu}F_{\mu\nu} + i \tilde{G}_\gamma
HF^{\mu\nu}\tilde{F}_{\mu\nu}\right]\,,\\ {\cal M}_{\gamma ZH} =
\fr{1}{v}\left[G_Z H Z^{\mu\nu}F_{\mu\nu} +i \tilde{G}_Z H
Z^{\mu\nu}\tilde{F}_{\mu\nu}\right]\,.
\end{array}\label{Lggam}
\ee
Here $F^{\mu\nu}$ and $Z^{\mu\nu}$ are the standard field
strengths for the electromagnetic and $Z$ field and $\tilde
{F}^{\mu\nu}=\varepsilon^{\mu\nu\alpha\beta}F_{\alpha\beta}/2$.
Dimensionless parameters $G_i$ are effective coupling constants.
They are sums of well known \SM contributions (see e.g.
\cite{anom1} for normalization)\fn{ With the proposed
experimental accuracy, when doing the final numerical
calculation, one should, of course, use $H\ggam$ coupling with
radiative corrections \cite{hdecay}.} and anomalous parts $g_i$
("anomalies"), describing the strength of interactions beyond
\SM, which are generally complex:
\be
G_i=G_i^{SM}+g_i\,,\quad
\tilde{G}_i=g_{Pi}\,;\quad g_a = |g_a|e^{i\xi_a}\,.
\label{LEFF}
 \ee

The complex values of "couplings" $g_a$ are quite natural. Indeed,
recall that even $G_i^{SM}$ are complex due to contributions, for
example, of $b$-quark loop in the amplitude. The same is valid in
various versions of the first variant of \CP violation. One
particular example of this is discussed in Sec.~5, where the
anomaly can be defined simply as the difference between the
minimal \SM and Two Doublet Higgs Model (II) with \CP violation.
If $\tan\beta\gg 1$, contribution of $b$ quarks in loops is
enhanced, which gives rise to the large imaginary part of the
amplitudes. For the second mechanism complex $g_i$ could be signal
of fundamental breaking of unitarity in theory.

We assume that future observations will reveal a picture close to
\SM\  and therefore anomalies $g_i$ will be small. In the first
mechanism of \CP violation  with $\Lambda\lsim M_H$ smallness of
anomalies is related to small values of corresponding mixing
angles $\alpha_m$, $g_i\sim\alpha_m$. In the second mechanism it
is related to large scale of New Physics $\Lambda$, i.e. $g_i=
(v/\Lambda_i)^2$ with $\Lambda_i\sim \Lambda$. The relation
between parameters $\Lambda_i$ and $\Lambda$ depends on the nature
of New Physics. \\
(A) The simplest extension of the \SM consists in adding new
charged heavy particles with mass $M_n$ that is not generated by
a Higgs mechanism (like in MSSM). They will circulate in loops
and give rise to anomalous effective $H\ggam$ and $HZ\gamma$
vertices, with $ \Lambda^2 \sim 4\pi M_n^2/\alpha$.\\
(B) If the heavy particle is a point-like Dirac monopole, then
$\Lambda^2 \sim \alpha M_n^2$.\\
(C) If New Physics is determined by higher dimension
(Kaluza--Klein) mechanism, the quantity $\Lambda$ is close
to the energy scale at which the extra dimensions come into
play.

For the second mechanism the anomalous amplitude is often
described with the aid of Effective Lagrangian with operators of
dimension 6, which has the same form as our effective vertices
(\ref{Lggam}). Our particular parameterization can be readily
linked to that used in other papers (e.g.
\cite{Il3,EfL1,hagiwara}). For example, correspondence of our
parameters $g_i$ to constants $d_i$ used in ref.~\cite{Il3}
reads $d_{Z\gamma}=  2g_{Z\gamma}/(c_Ws_W)$, $\bar{d}_{Z\gamma}=
g_{PZ}/(c_Ws_W)$.

Finally, we undertake a study where both $|g_i|$ and $\xi_i$ are
treated  as {\em independent parameters}. This is done in
contrast to other similar investigations, where the complexity
was not an explicitly free parameter, but fixed by the
particular model considered. We argue that our approach accounts
for the most wide range of possible anomalies. Determination of
both sets of parameters should be considered primarily as an
experimental task\fn{ Certainly, only phase differences are
measurable for entire effective couplings. Expecting relatively
small magnitude of anomaly, one can conclude that the phases of
entire quantities $G_\gamma$, $G_Z$ are close to their \SM
values $\xi_\gamma^{SM}$ and $\xi_Z^{SM}$ and the effect of
anomaly itself is reduced by factor $\cos(\xi_\gamma-
\xi_\gamma^{SM}) $.}.\\

{\bf About figures and notation there}. Currently, due to the
large number of new model parameters, a thorough investigation
of regions of the parameter space, achievable in future
experiments, makes little sense. Instead of that we present in
our figures examples for some values of parameters, which
illustrate that the study of these effects at the Photon
Colliders is indeed possible and profitable.

There are no doubts that relatively large anomalies will be
discovered easily. Therefore, we concentrate our efforts on the
case when the anomalous effects are relatively small as compared
with basic \SM effects. In this case the effects of anomalies will
be seen mainly in the interference with the \SM effects, and
contributions of different anomalies in the observed cross
sections are additive with good accuracy. This is why we treat
each anomaly separately, assuming all other anomalies absent
(the corresponding $g_i=0$).

\section{\bm Process $\ggam\to H$}

Let us denote by $\langle\sigma^{SM}\rangle_{np}$ the \SM Higgs boson
production cross section in unpolarized photon collisions averaged
over a certain small effective mass interval (see e.g. \cite{anom1}).
Then the cross section of the Higgs boson production
can be written in the form:
\be\begin{array}{c}
\langle\sigma\rangle(\lambda_i,\ell_i,\psi)=\langle\sigma^{SM}\rangle_{np}
 T(\lambda_i,\ell_i,\psi)\,;\\[3mm]
 T(\lambda_i,\ell_i,\psi)=\fr{|G_\gamma|^2}{|G_\gamma^{SM}|^2}
 (1+\lambda_1\lambda_2
 +\ell_1\ell_2 \cos 2\psi)
 +\fr{|\tilde{G}_\gamma|^2}{|G_\gamma^{SM}|^2}
 \left(1+\lambda_1\lambda_2-\ell_1\ell_2 \cos 2\psi \right)\\ +
 2\fr{Re(G_\gamma^*\tilde{G}_\gamma)}{|G_\gamma^{SM}|^2}(\lambda_1+\lambda_2)
 +2\fr{Im(G_\gamma^*\tilde{G}_\gamma)}{|G_\gamma^{SM}|^2}
 \ell_1\ell_2\sin 2\psi\,.
 \end{array}\label{poldep}
 \ee
Here $\lambda_i$ and $\ell_i$ ($i=1,2$) are degrees of circular
and linear polarization respectively of the photon beams and
$\psi$ is the polar angle between the linear polarization vectors
of the two photon beams.

In the \SM case we have only the first item in this sum. (Note that
the $\ggam\to b\bar{b}$ background is practically independent on
linear polarization of photons.)

An important feature here is interference terms. They give rise to
the inequality of the two directions of rotation and to the
modification of the $\psi$-dependence, which is entirely due to the
\CP odd admixture to \CP even Lagrangian. Owing to these
modifications, a number of experimentally measurable quantities
appear that can help study \CP even and odd anomalies separately.

It is useful to introduce five different asymmetries:
\bear{c}
T_\pm=\fr{\langle\sigma\rangle(\lambda_i,\ell_i=0)\pm
\langle\sigma\rangle(-\lambda_i,\ell_i=0)}
 { 2\langle\sigma^{SM}\rangle_{np}}
 \propto\left\{
 \begin{array}{l}
 (1+\lambda_1\lambda_2)(|\tilde{G}_\gamma|^2+|G_\gamma|^2)\,,\\
  2(\lambda_1+\lambda_2)Re(\tilde{G}^*_\gamma
 G_\gamma)\,;
 \end{array} \right.\\[4mm]
T_{\|}=\fr{\langle\sigma\rangle(\lambda_i=0,\ell_i,\psi=0)}
{\langle\sigma^{SM}\rangle_{np}}
\propto\left[|G_\gamma|^2(1+\ell_1\ell_2)
 +|\tilde{G}_\gamma|^2(1- \ell_1\ell_2)\right]\,,\\[4mm]
 T_\bot=\fr{\langle\sigma\rangle(\lambda_i=0,\ell_i,\psi=\pi/2)}
 {\langle\sigma^{SM}\rangle_{np}}
 \propto\left[|G_\gamma|^2(1-\ell_1\ell_2)
 +|\tilde{G}_\gamma|^2(1+ \ell_1\ell_2)\right]\,,\\[4mm]
 T_\psi=\fr{\langle\sigma\rangle(\lambda_i=0,\ell_i,\psi =3\pi/4)
 -
 \langle\sigma\rangle(\lambda_i=0,\ell_i,\psi =\pi/4)}
 {\langle\sigma^{SM}\rangle_{np}}
 \propto 2\ell_1\ell_2Im(\tilde{G}^*_\gamma G_\gamma)\,,\\[5mm]
 \end{array}\label{somepoldep2}
 \ee
whose \SM values are
$$
T_+^{SM} = 1 +\lambda_1\lambda_2,\;\;
T_-^{SM} = 0,\;\; T_{\|}^{SM} = 1 + \ell_1\ell_2,\;\; T_{\bot}^{SM} =
1 - \ell_1\ell_2,\;\; T_{\psi}^{SM} = 0.
$$

\begin{figure}[!tbh]
 \begin{center}
 \includegraphics[width=0.43\textwidth]{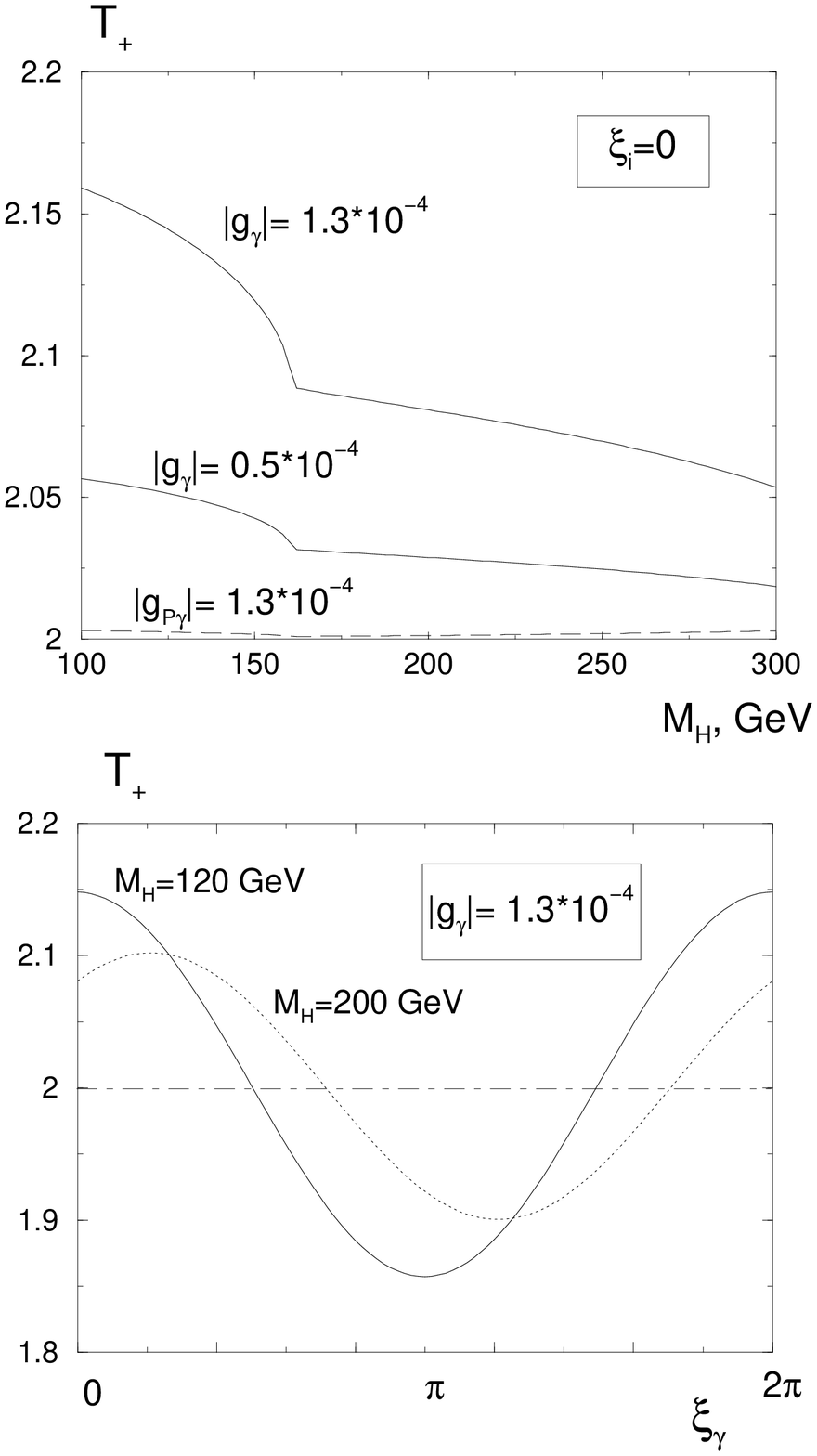}
 \hfill
 \includegraphics[width=0.46\textwidth]{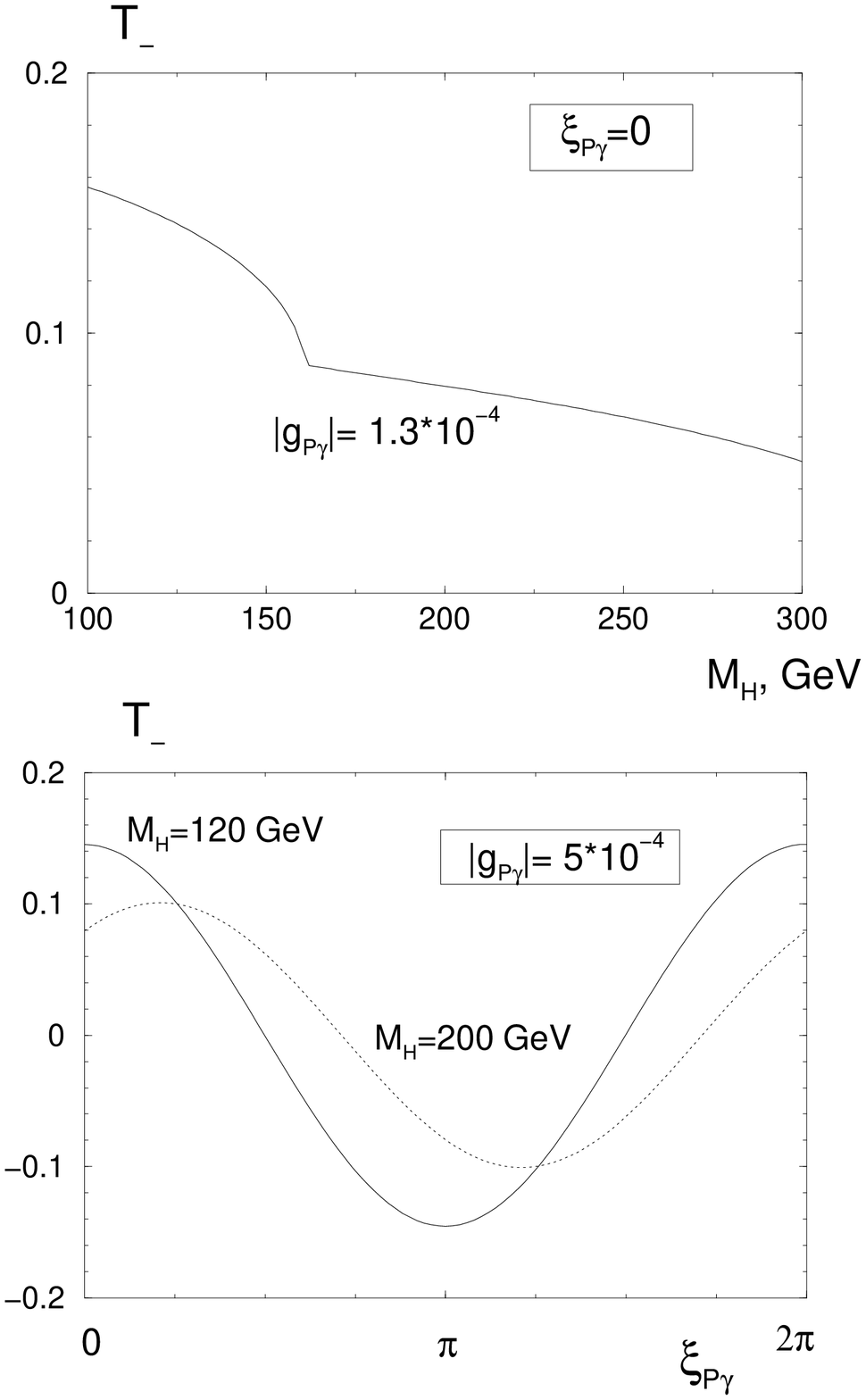}
 \\
 \parbox[t]{0.43\textwidth}
 {\caption{\em The quantity $T_+$; $\lambda_1\lambda_2 =1$}
 \label{figggh1}}
 \hfill
 \parbox[t]{0.46\textwidth}
 {\caption{\em The quantity $T_-$; $\lambda_1\lambda_2 =1$}
 \label{figggh2}}
 \end{center}
 \end{figure}

\begin{figure}[!tbh]
 \includegraphics[width=0.43\textwidth]{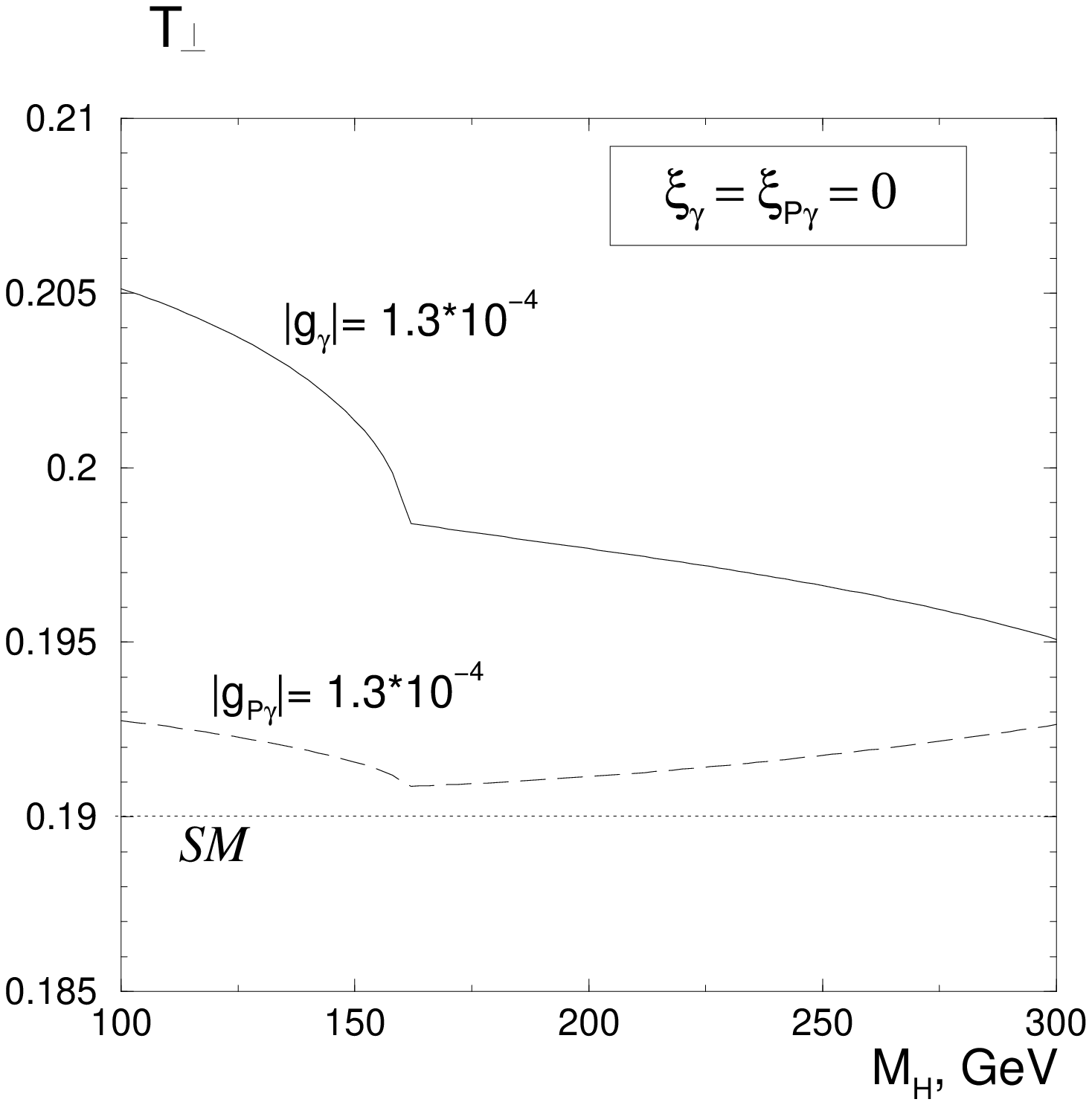}
 \hfill
 \includegraphics[width=0.48\textwidth]{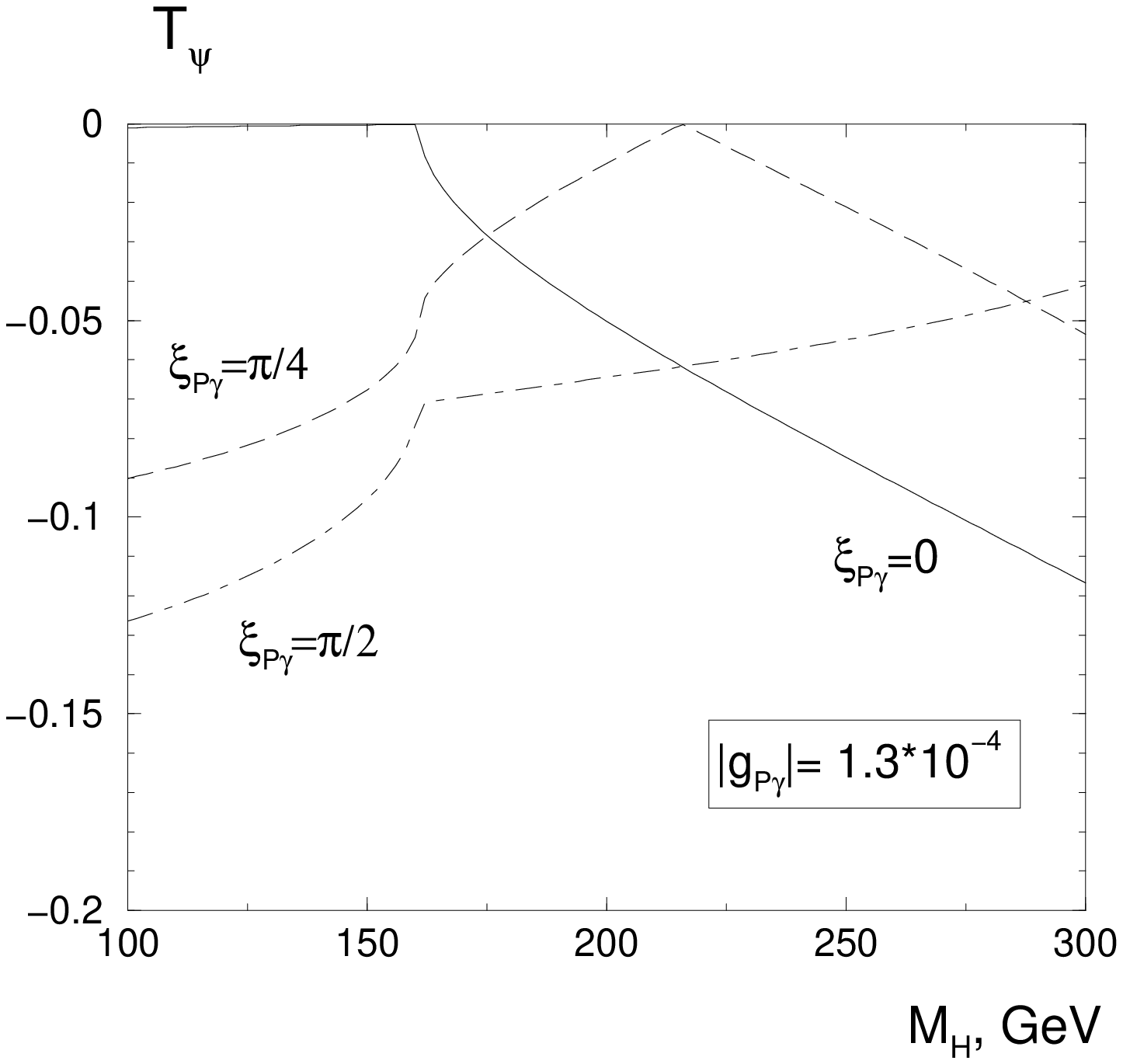}
 \\
 \parbox[t]{0.43\textwidth}
 {\caption{\em The quantity $T_\bot$; $\ell_i =0.9$}
 \label{figggh3}}
 \hfill
 \parbox[t]{0.48\textwidth}
 {\caption{\em The quantity $T_\psi$; $\ell_i =0.9$}
 \label{figggh4}}
 \end{figure}

The quantities $T_+$, $T_\|$ and $T_\perp$ are combinations of
$|{G}_\gamma|^2$ and $|\tilde{G}_\gamma|^2$ with different
weights.  These asymmetries are sensitive to the \CP even anomaly
and its phase $\xi_\gamma$ via its interference with the \SM
contribution. The best quantity for this study is of course $T_+$,
which is illustrated by Fig.~\ref{figggh1}.  Certainly, curves for
\CP even anomaly effects at $\xi_\gamma=0$ are the same as
obtained in Ref.~\cite{anom1} (modulo to reparameterization
of anomalous terms).
These three quantities include also the \CP odd anomaly in the
form $|\tilde{G}_\gamma|^2$, which is $\sim g^2_{P\gamma}$,
i.e. small and independent of $\xi_{P\gamma}$
(the corresponding $g_{P\gamma}$ dependence
was studied in ref.~\cite{Goun}). Even in the case of $T_\perp$,
where the contribution of $|\tilde{G}_\gamma|^2$ is enhanced, it
is difficult to see the effect of \CP-odd anomalies at reasonably
small $g_{P\gamma}$, Fig.\ref{figggh3}.

 The remaining two quantities --- $T_-$ and $T_\psi$ --- are much more
 useful for study of \CP violating effects in $\ggam H$ interaction.
 Their study supplements each other.  Both of them differ from zero
 only if the \CP parity is broken. They derive from the interference of
 the \CP odd and \CP even items in (\ref{Lggam}).
 Fig.~\ref{figggh2} shows the $T_-$ dependence on
 $|g_{P\gamma}|$ and phase $\xi_{P\gamma}$ for different values of
 the Higgs boson mass. At $M_H<160$ GeV ($WW$ threshold) the basic
 quantity $G_\gamma^{SM}$ is practically real.  Therefore, the
 quantity $T_-$ has maximum at $\xi_{P\gamma}=0$.  Above this
 threshold the imaginary part of $G_\gamma^{SM}$ becomes substantial,
 and the position of maximum is shifted to $\xi_{P\gamma} \neq 0$.
 Fig.~\ref{figggh4} shows that the \CP odd anomaly effect is strong
 in this asymmetry as well, and exhibits a remarkable dependence
of $T_\psi$ on the value of phase $\xi_{P\gamma}$.
With measurement of $T_-$ and $T_\psi$ one can
 extract from the data both $|g_{P\gamma}|$ and $\xi_{P\gamma}$
 since $T_-$ and $T_\psi$ represent the real and imaginary part of
the same quantity.

\section{\bm Process \egeh}

The process \egeh\ is considered here as a good tool for study of
$HZ\gamma$ anomalous interactions provided $H\ggam$ anomalies are
known from the experiments in the \ggam\ mode.
This process was studied within \SM in detail
in refs.~\cite{anom1,Il1}\fn{ The production of the
pseudoscalar Higgs boson in such a reaction was studied e.g. in
ref.~\cite{savci}, see also ref.~\cite{CDT} for the \MSSM\ case.}.
It is described by diagrams of three types --- those with photon
exchange in $t$-channel, with $Z$ exchange in $t$ channel and box
diagrams.  This subdivision is approximately gauge invariant with
accuracy $\sim m_e/M_Z$ \cite{anom1}. The difference in the cross
sections $\sigma^L$ and $\sigma^R$ for the left-hand and
right-hand polarized electrons is obliged to interference between
photon and $Z$ exchange amplitudes.

\begin{figure}
 \begin{center}
 \includegraphics[height=8cm]{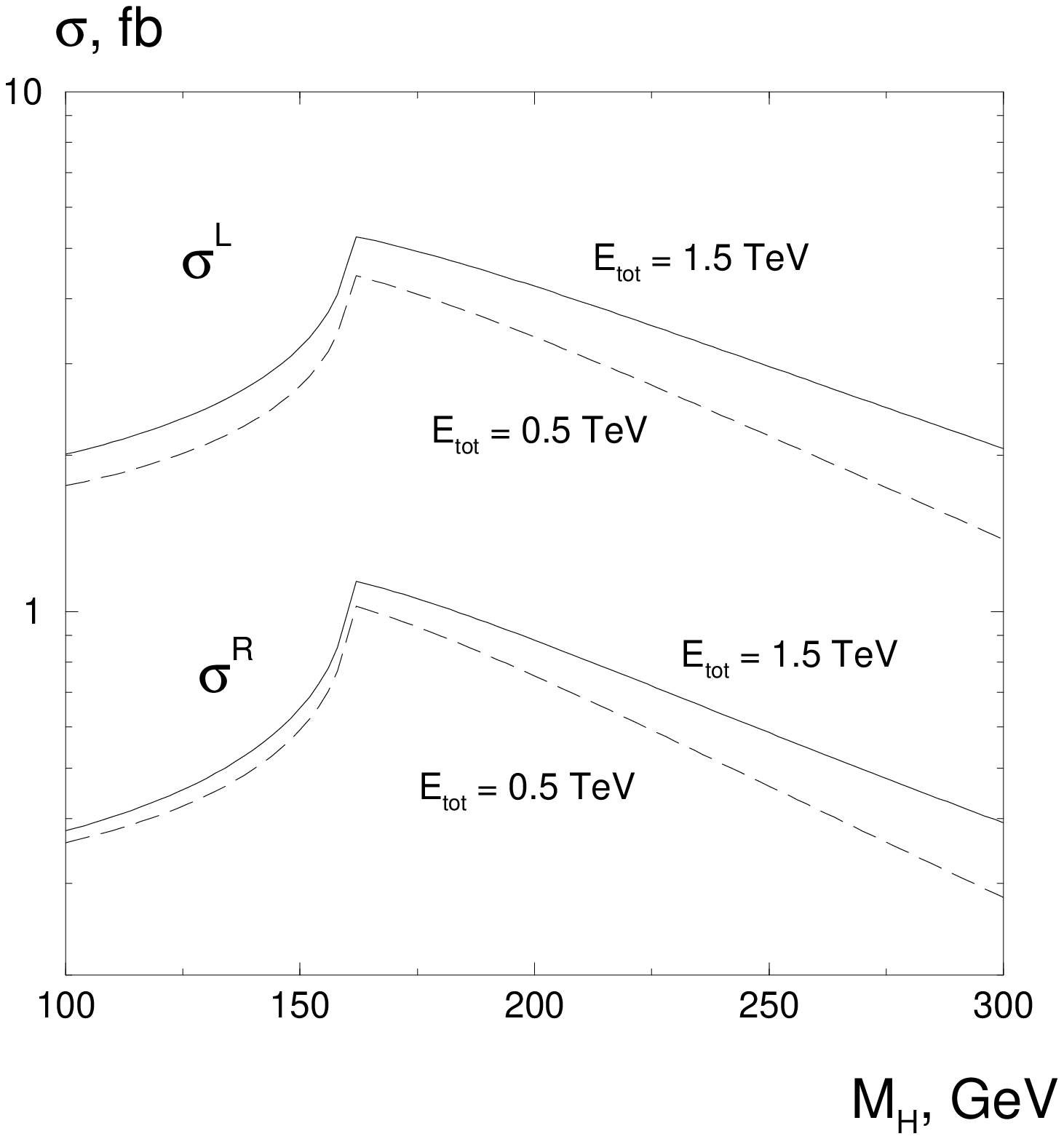}
 \caption{\em The \SM cross section of \egeh process,
 $Q^2>1000$ GeV$^2$.}
 \label{figegehsm}
 \end{center}
 \end{figure}

The main contribution into the total cross section is given by
diagrams with photon exchange in $t$-channel.  Therefore, this total
cross section is sensitive to the $H\ggam$ anomalies and weakly
sensitive to the $HZ\gamma$ anomalies, which are our major concern
here (the difference $\sigma^L-\sigma^R$ is small as compared with the
unpolarized cross section).  This picture is improved with the growth of
transverse momentum of the scattered electron $p_\bot$. Indeed, with this
growth photon exchange contribution is strongly reduced, while
$Z$--boson exchange contribution changes only marginally at
$p_\bot\lsim M_Z$.  At transverse momenta of the scattered electrons
$p_\bot>30$ GeV and for longitudinally polarized initial electrons
the effect of Z-exchange should be seen well \cite{anom1}. To feel
the scale of observed effects, we present in Fig.~\ref{figegehsm} the
\SM cross sections $\sigma^L$ and $\sigma^R$
integrated over the region $Q^2>1000$ GeV$^2$ and averaged over
initial photon polarizations.  We use this limitation in $Q^2$
everywhere below.

We denote the particle momenta as $p$ for the incident electron, $k$
for the photon, $p'=p-q$ for the scattered electron and $Q^2=-q^2$.
In our calculations far from the photon pole in $t$-channel we neglect
the electron mass. We also denote: $u=2kp'=M_H^2+Q^2-s$, $x=2kq/s
\equiv (M_H^2+Q^2)/s$, $E_{tot}=\sqrt{s}$. The collision axis
is labeled as $z$ axis and $x$ axis is chosen along the direction of
the photon linear polarization vector $\vec{\bf{\ell}}$.  Finally,
angle $\phi$ is the azimuthal angle of the scattered electron
relative to so-defined $x$ axis.  The values $\zeta=-1$ or $\zeta=+1$
correspond to left-hand or right-hand polarized initial electrons.
We use superscripts $L$ and $R$ to label quantities referring to
these polarizations.\\

{\bf The qualitative features of the observable effect} could be
understood taking into account that the quantities below could be
treated as the average helicity $\lambda_V$ and degree of linear
polarization $\ell_V$ of an exchanged virtual photon or $Z$ boson:
\be
 \lambda_V=\fr{s^2-u^2}{s^2+u^2}\zeta=\fr{x-x^2/2}{1-x+x^2/2}\zeta\,,\quad
 \ell_V=\fr{2s|u|}{s^2+u^2}=\fr{1-x}{1-x+x^2/2}\,,\label{virtpol}
 \ee
with vector of linear polarization $\vec{\bf{\ell}}_V$ lying in
the electron scattering plane \cite{BGMS}. Since usually $x\ll 1$,
we have $\lambda_V\ll 1$ and $\ell_V\approx 1$. Therefore, joining
the results of the previous section and those from
ref.\cite{anom1}, one can conclude that the effect of \CP odd
$HZ\gamma$ interaction can be seen in the dependence on angle
$\phi$ in the experiments with left-- and right-- polarized
electrons and in the study of dependence on the sign of the
incident photon helicity. These dependencies have not been studied
earlier.\\

{\bf Helicity amplitudes} of the process are calculated just as in
Ref.~\cite{anom1}. With notations for the box contributions from that
paper we have (in these equations helicities $\lambda\,,\;
\zeta=\pm 1$)
\bear{c}
{\cal M}=-\fr{4\pi\alpha}{M_Ws_W}\sqrt{\fr{Q^2}{2}}\left\{
s\fr{1+\zeta\lambda}{2}+
(s-M_H^2-Q^2)\left[\fr{1-\zeta\lambda}{2}\cos 2\phi
+\fr{\zeta-\lambda}{2} i\sin 2\phi\right]\right\}\\ \times(\lambda K
+\tilde{K})\,,\quad \left(K=V-\zeta A+B_+\,,\;\; \tilde{K}=
\tilde{V}-\zeta\tilde{A}+\zeta B_-\right)\,.
 \end{array}\label{egehampl}
 \ee
Here $V$ and $A$ stand for vector and axial $t$--channel exchange
contributions, $B_\pm$ are the box contributions which are composed
from items related to the $W$ or $Z$ circulating in box\fn{ The
box diagrams contribution (and their interference with other
diagrams) is small in comparison with other contributions.}:
\bear{c}
V=\fr{G_\gamma}{Q^2} + \fr{v_e G_Z}{4s_Wc_W(Q^2+M_Z^2)}\,,\quad
A=-\fr{G_Z}{4s_Wc_W(Q^2+M_Z^2)}\,,\\[5mm]
\tilde{V}=\fr{\tilde{G}_\gamma}{Q^2} +
 \fr{v_e \tilde{G}_Z}{4s_Wc_W(Q^2+M_Z^2)}\,,\quad
 \tilde{A}=-\fr{\tilde{G}_Z}{4s_Wc_W(Q^2+M_Z^2)}\,;\\[5mm]
 B_\pm=\fr{\alpha M_W^2}{4\pi s_W^2}\cdot
 \left[\fr{W(s,u) \pm W(u,s)}{2} + \fr{Z(s,u)
 \pm Z(u,s)}{2}\right]\,.
 \end{array}\label{vanot}
 \ee

The amplitude squared for an arbitrarily polarized photon beam can be
written in terms of helicity amplitudes and the photon density matrix
$\rho$ written in helicity basis as
\be
|{\cal M}|^2 = {\cal M}^*_a\rho_{a\,b}{\cal M}_b, \;\;
a,b=+,-\,,\quad \rho=\fr{1}{2}\left(\begin{array}{ccc} 1+\lambda
&\;\;&-\ell\\ -\ell&&1-\lambda \end{array}\right)\,.
\ee
So that the cross section reads (here $\zeta=\pm 1$):
\bear{rl}
d\sigma =&
\fr{\pi\alpha^2}{2M_W^2s_W^2}\fr{d\phi}{2\pi}Q^2dQ^2\fr{s^2+u^2}{2s^2}
 \left(U_0 + \lambda U_\lambda + \ell\cos 2\phi\ U_\bot -\ell\sin
 2\phi\ U_\psi \right)\,;\\[3mm] &U_0 =
 \left(|K|^2+|\tilde{K}|^2\right)+ \lambda_V
 2Re\left(K\tilde{K}^*\right),\;\;U_\bot  =
 \ell_V\left(|K|^2-|\tilde{K}|^2\right)\,\\
&U_\lambda =
 2Re\left(K\tilde{K}^*\right) +
 \lambda_V\left(|K|^2+|\tilde{K}|^2\right),\;\;
 U_\psi = 2 Im\left(K\tilde{K}^*\right)\,.
 \end{array}\label{egehpol}
 \ee

With notations (\ref{virtpol}) it becomes evident that this equation
reproduces term by term the polarization dependencies of $\ggam\to H$
process (\ref{poldep}), in particular, $T_+,\,T_\|\to U_0$,
$T_\bot\to U_\bot$, $T_-\to U_\lambda$, $T_\psi\to U_\psi$.
Therefore, the similar studies of $HZ\gamma$ interaction are possible
here.  However, there is a difference between effects of linear
photon polarization in these two reactions. In the \ggam\ collisions
we can control linear polarizations and {\em choose} their relative
orientation to study specific contribution. In the \egam\ collision
we cannot control relative orientation of linear polarizations, so
that some Fourier-type analysis is necessary to see contributions
under interest.\\

{\bf Different asymmetries}.  The quantities $U_0$ and $U_\bot$
are weakly sensitive to the $\tilde{G}_Z$.The
sensitivity of $U_0$ to the \CP even anomalous interaction was studied, in
fact, in refs.~\cite{anom1,Il3}.

  \begin{figure}[!tbh]
 \includegraphics[width=0.42\textwidth]{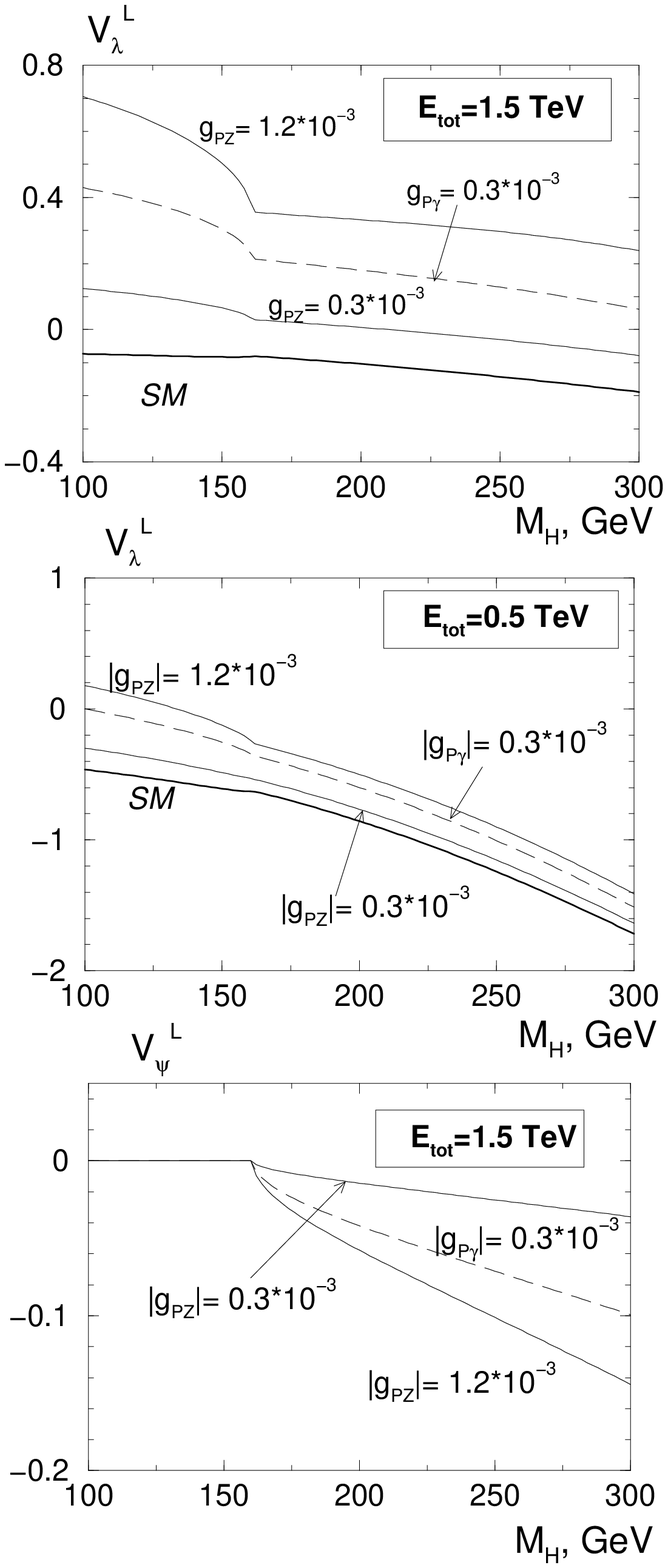}
 \hfill
 \includegraphics[width=0.48\textwidth]{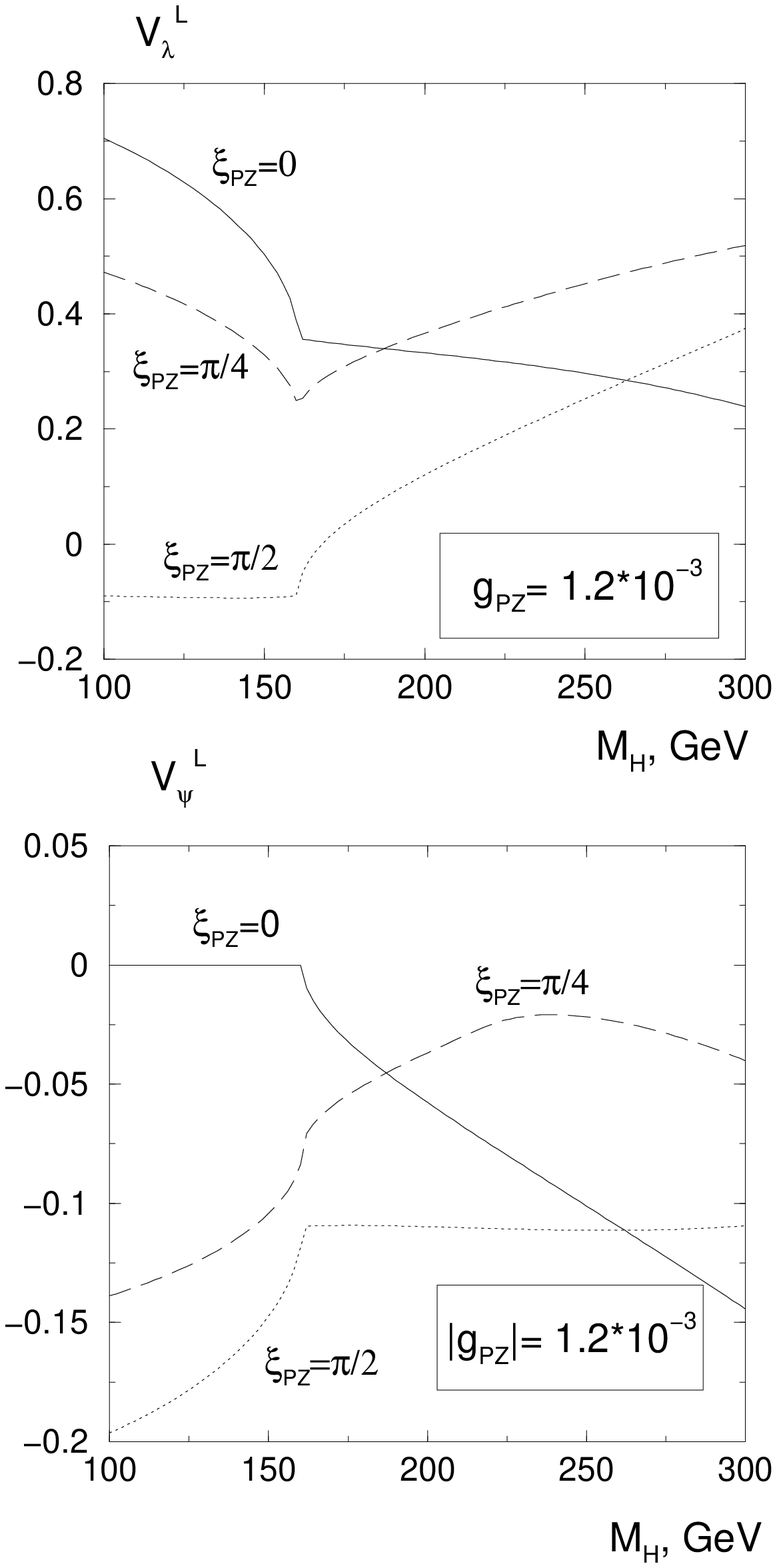}
 \\
 \parbox[t]{0.42\textwidth}
 {\caption{\em The asymmetries $V_\lambda$ and $V_\psi$;
 $Q^2>1000$ GeV$^2$, $\xi_{P\gamma}=\xi_{PZ}=0$}
 \label{figegeh1}}
 \hfill
 \parbox[t]{0.48\textwidth}
 {\caption{\em The asymmetries $V_\lambda^L$ and $V_\psi$
 at different $\xi_{Pi}$, $Q^2>1000$ GeV$^2$}
 \label{figegeh2}}
 \end{figure}

The quantities $U_\lambda$ and $U_\psi$ are most sensitive to the \CP
odd anomalies. Thus, we consider asymmetries
\bear{c}
V_\lambda^{L,R}=\fr{\int d\sigma^{L,R}(\lambda)- \int
d\sigma^{L,R}(-\lambda)}{|\lambda|\int d\sigma^{SM}_{np}} \propto\int
U_\lambda^{L,R} \,,\\
[4mm]
V_\psi^{L,R} =\fr{\int d\sigma^{L,R} \sin
2\phi}{|\ell|\int d\sigma^{SM}_{np}} \propto\int U_\psi^{L,R}\,,
\end{array}\label{egammeas}
\ee
with integrations spanning over the region $Q^2>Q^2_0=1000$ GeV$^2$
and the whole region of $\phi$ for the left--hand and right--hand
polarized initial electrons. (The integrals in denominators are
calculated for the nonpolarized initial particles.) It happens that
the cross sections for the left-hand polarized electrons are much
higher than those for the right-handed electrons (see
Fig.~\ref{figegehsm}). Therefore, we present graphs for the
left-handed electrons only. The anomalous effect for the right-handed
electrons is also small in its absolute value. We have not
encountered any case where $\sigma^R$ would be a useful source of
additional information, despite the relative value of anomaly
contribution can be higher here.

{\bf\bm The quantity $V_\lambda^L$} describing the helicity asymmetry
is analogous to $T_-$ in the \ggam\ case with accuracy to
contribution $\sim(|K|^2+|\bar{K}|^2)$ entering with small
coefficient $\lambda_V$.  This contribution results in non-zero
$V_\lambda$ even in \SM.  Figs.~\ref{figegeh1} shows dependence of
this quantity on $|g_{PZ}|$.  For the purposes of comparison, the
effect of a $g_{P\gamma} = 0.3\cdot 10^{-3}$ $H\ggam$ anomaly is also
shown. We see that the values of this helicity asymmetry are large
enough.  Note that the signal/background ratio improves with the
growth of energy since the \SM contribution into the discussed
quantity decreases approximately $\propto \lambda_V\sim s^{-1}$ while
the anomaly effect increases weakly, $\propto\ln(s/M_Z^2)$.

The same figure depicts also {\bf\bm the quantity $V^L_\psi$} at
different values of $|g_{PZ}|$. Again we also draw a comparison
with a $H\ggam$ \CP-odd anomaly. This quantity is intrinsically
smaller than $V^L_\lambda$, so the \CP-odd $HZ\gamma$ anomaly can be
seen only at $|g_{PZ}| > 10^{-3}$.

The dependence of $V_\lambda$ and $V_\psi$ on the phase of $HZ\gamma$
anomaly $\xi_{PZ}$ is shown in Fig.~\ref{figegeh2}.  (The dependence
of these quantities on the parameters of $H\ggam$ anomaly has the
same form but the magnitude is somewhat larger.) These curves
closely resemble dependencies of $T_-$ and $T_\psi$ on $\xi_{P\gamma}$
in the $\ggam\to H$ case. We see the familiar phase dependence
$\propto \cos(\xi_{PZ}-\bar{\xi}_\gamma)$ or $\sin(\xi_{PZ}-
\bar{\xi}_\gamma)$ (here $\bar{\xi}_i$ are phases of $G_\gamma$ and
$G_Z$ which are close to their \SM values).  The effect of switching
on of the imaginary part of the \SM contribution at $M_H\sim 160$ GeV is
clearly seen in these curves. In the phenomenological analysis, it is
helpful that $V_\lambda$ and $V_\psi$ are intrinsically
complementary: just as it was in $\ggam\to H$ case, $V_\lambda$ is
the real part and $V_\psi$ is the imaginary part of the same
quantity. Therefore, at any value of $M_H$ and $\xi_{PZ}$ either
$V_\lambda$ or $V_\psi$ will deviate strongly from the \SM value.

\section{ Scalar-pseudoscalar mixing within two doublet Higgs
model}\label{secmix}

A specific case of \CP violation takes place in the scalar--axial
mixing within the two doublet Higgs model (2HDM).  This model is
described with the aid of mixing angle $\beta$ (defined via the ratio
of v.e.v.'s for two basic scalar fields, $\tan\beta=
\langle\phi_1\rangle/\langle\phi_2\rangle $) and three Euler
mixing angles $\alpha_1$, $\alpha_2$, $\alpha_3$ (see, for example,
ref.~\cite{ggk}). The observed neutral Higgs bosons are combined from
the basic scalar fields as
\bear{c}
\left(\begin{array}{c}h_1\\h_2\\h_3\end{array}\right)
 =-\sqrt{2}R\left(\begin{array}{c}Re\phi_1^0\\
 Re\phi_2^0\\ Im \left(s_\beta
 \phi_1^0 -c_\beta \phi_2^0\right)\end{array}\right),
 \\[3mm]
 R= \left(\begin{array}{ccc} c_1&
 -s_1c_2 &s_1s_2 \\ s_1c_3
 &\;c_1c_2c_3-s_2s_3\;&-c_1s_2c_3-c_2s_3\\
 s_1s_3&c_1c_2s_3+s_2c_3&-c_1s_2s_3+c_2c_3
 \end{array}\right)\,
 \end{array}\label{mixing}
 \ee
Here $c_i=\cos\alpha_i,\;\;s_i=\sin\alpha_i$. Our definition differs
from that used in Ref.~\cite{ggk} by the sign minus in front of $R$
in (\ref{mixing}). The \CP conserving case is realized at
$\alpha_2=\alpha_3=0$, the last angle $\alpha_1$ is related to the
quantity $\alpha$ used for the case without \CP violation as
$\alpha_1 \to \pi/2 - \alpha$, $h_1\to h$, $h_2\to A$, $h_3\to -H$.
Instead of $\alpha_1$, we use below the angle $\delta =\beta-(\pi/2
-\alpha_1)$.

We consider only the lightest Higgs boson $h_1$ having in mind the
decoupling regime where $M_{H^\pm},\,M_{h_2},\,M_{h_3}\gg
M_{h_1}$. Besides, we fix the only relevant free parameter of 2HDM
in the Higgs self-interaction as $\lambda_5= 2M^2_{H^\pm}/v^2+g^2$
(just as it is in MSSM, see \cite{hunter} for definition). This
choice guarantees us negligibly small contribution of charged
Higgs loops into the discussed couplings of Higgs boson with photons.

To describe couplings of the lightest Higgs boson $h_1$ with quarks
and charged leptons we use the widespread "Model II'' in which the
ratios of these couplings to those in the minimal \SM\ (one Higgs
doublet) are
\bear{c}
 \bar{u}h_1u\to
(\sin\delta+\cot\beta\cos\delta)\cos\alpha_2-
i\gamma^5\cot\beta\cos(\delta-\beta)\sin\alpha_2\,,\\ [4mm]
\bar{d}h_1d,\,\bar{\ell}h_1\ell\to
 (\sin\delta-\tan\beta\cos\delta)-i\gamma^5\tan\beta\cos(\delta-\beta)
 \sin\alpha_2\,,\\
 [4mm]
 VVh_1\to
 \sin\delta-\sin\beta\cos(\delta-\beta)(1-\cos\alpha_2)\,.
 \end{array}\label{mixcoupl1}
 \ee

The effective couplings of Higgs boson with light $G_i$
(\ref{Lggam}) can be written via standard loop integrals
 and the above mixing angles (see \cite{anom1} for definitions).
\bear{c}
G^\gamma=G^\gamma_{SM}\sin\delta
+\fr{\alpha}{12\pi}\cos\delta\left[-\Phi_{1/2}(b)\tan\beta +
4\Phi_{1/2}(t)\cot\beta\right] + \mbox{\em scalars}\\ [4mm] -
\fr{\alpha}{12\pi}(1-\cos\alpha_2) \left[
 3\Phi^\gamma_1(W)\sin\beta\cos(\delta-\beta)
 +4\Phi_{1/2}(t)(\sin\delta+\cot\beta\cos\delta) \right]\,,\\
 [4mm]
 \tilde{G}^\gamma=\fr{\alpha}{12\pi}\left[\Phi_{1/2}^A(b)\tan\beta
 +4\Phi_{1/2}^A(t)\cot\beta\right]\cos(\delta-\beta)\sin\alpha_2\,,\\
 [4mm]
 G^Z=G^Z_{SM}\sin\delta
 +\fr{\alpha}{4\pi}\left[v_b\Phi_{1/2}(b)\tan\beta
 +2v_t\Phi_{1/2}\cot\beta \right]\cos\delta + \mbox{\em
 scalars}\\
 [4mm]
 -\fr{\alpha}{4\pi}(1-\cos\alpha_2)
 \left[\Phi_1^Z(W)\sin\beta\cos(\delta-\beta)+2v_t\Phi_{1/2}(t)
 (\sin\gamma+\cot\beta\cos\delta)
 \right]\,,\\
 [4mm]
 \tilde{G}^Z =
 \fr{\alpha}{4\pi}\left[2v_t\Phi^A_{1/2}(t)\cot\beta
 -v_b\Phi^A_{1/2}(b)\tan\beta\right]\cos(\delta-\beta)\sin\alpha_2\,;\\
 v_b=-\fr{3-4s_w^2}{12s_wc_w}\,,\quad
 v_t=\fr{3-8s_w^2}{12s_wc_w}\,
 \end{array}\label{phigcp}
 \ee
The first lines in formulas for
$G_\gamma$ and $G_Z$ give their form for the standard 2-doublet
model without \CP-mixing. At large $\tan\beta$ the imaginary part of
all these couplings (arising from the $b$-quark contribution) becomes
essential. It gives phases $\xi_i$ (\ref{LEFF}) which differ
essentially from 0 or $\pi$.
The corresponding values of  $g_i$ and phases
$\xi_i$ (\ref{LEFF}) could be calculated easily from
these equations. The word {\em scalars} means charged Higgs loop
contribution, it is negligibly small in the discussed case, so we
will not write it below.

Finally, all box diagrams include $VVh$ vertex. Therefore the box
contribution (\ref{vanot}) to the amplitude changes as
\be
B_\pm\to
B_\pm^{SM}\left[\sin\delta-\sin\beta\cos(\delta-\beta)(1-\cos\alpha_2)
\right]\,.
\ee

 \begin{figure}[!tbh]
 \begin{center}
 \includegraphics[height=8cm]{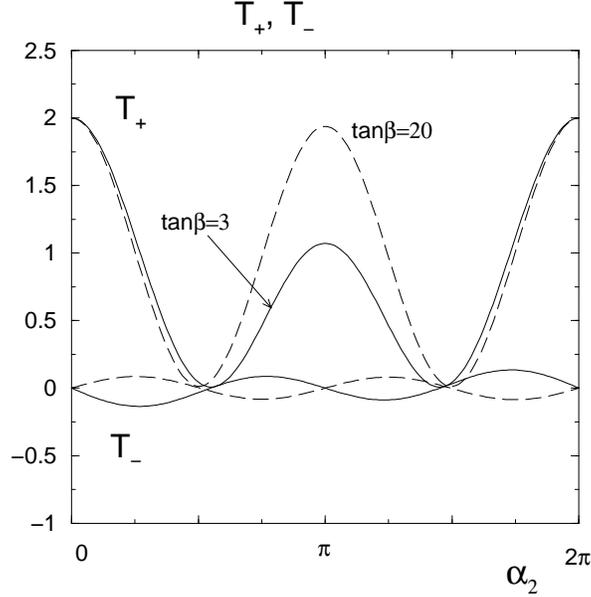}
 \caption{\em Quantities $T_\pm$ in 2HDM(II), $M_h=110$ GeV.
Strong \CP violation} \label{figmixgghscan}
 \end{center}
 \end{figure}

To make new effects more manifest, we study the dependence on two
parameters $\alpha_2$ and $\beta$ only, keeping main features of
discussed Higgs boson $h_1$ as close as possible to the Higgs boson of
\SM. For this purpose we fix parameter $\delta\approx\pi/2$
and consider small enough values of \CP--violated mixing angle
$\alpha_2$. According to eq.~(\ref{mixcoupl1}), in this case
couplings of $h$ with quarks and gauge bosons are close to those in
\SM (see refs.~\cite{gko} for the detail discussion of this
opportunity). In this case we have instead of previous equations
\bear{c}
\bar{u}h_1u\to \cos\alpha_2-
i\gamma^5\cos\beta\sin\alpha_2\,,\;\; \bar{d}h_1d\to 1-
i\gamma^5\tan\beta\sin\beta\sin\alpha_2\,,\\
[4mm]
VVh_1\to
1-\sin^2\beta(1-\cos\alpha_2)\,.  \end{array} \ee \bear{c} G^\gamma
=G^\gamma_{SM} - \fr{\alpha}{12\pi}(1-\cos\alpha_2) \left[
3\Phi^\gamma_1(W)\sin^2\beta +4\Phi_{1/2}(t) \right]\,,\\
[4mm]
\tilde{G}^\gamma=\fr{\alpha}{12\pi}\left[\Phi_{1/2}^A(b)\tan\beta
 +4\Phi_{1/2}^A(t)\cot\beta\right]\sin\beta\sin\alpha_2\,,\\
 [4mm]
 G^Z=G^Z_{SM}
 -\fr{\alpha}{4\pi}(1-\cos\alpha_2)
 \left[\Phi_1^Z(W)\sin^2\beta+2v_t\Phi_{1/2}(t)
 \right]\,,\\
 [4mm]
 \tilde{G}^Z =
 \fr{\alpha}{4\pi}\left[2v_t\Phi^A_{1/2}(t)\cot\beta
 -v_b\Phi^A_{1/2}(b)\tan\beta\right]\sin\beta\sin\alpha_2\,,\\
 [4mm]
 B_\pm =B_\pm^{SM}[1-\sin\beta\cos\beta(1-\cos\alpha_2)]\,.
 \end{array}\label{CPMSM}
 \ee

 \begin{figure}[!tbh]
 \begin{center}
 \includegraphics[height=10cm]{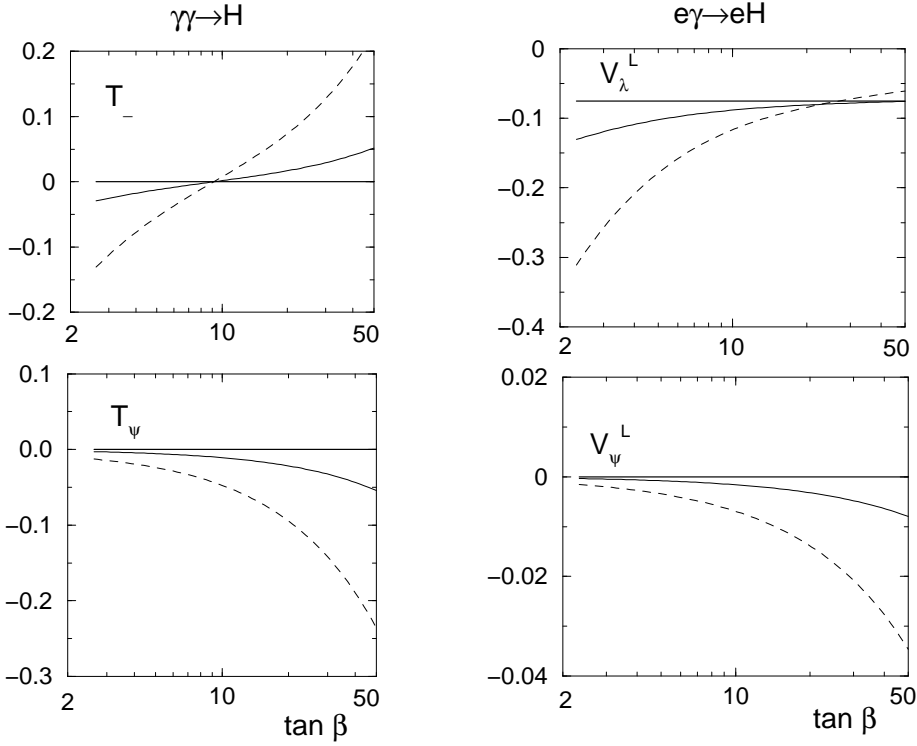}
 \caption{\em Spin asymmetries in $\ggam\to h$ and \egeh
processes due to scalar-pseudoscalar mixing in 2HDM(II), $M_h=110$
GeV; for \egeh process $E_{tot}=1.5$ TeV. The thick solid lines show
the \SM values; thin solid and dashed lines refer to $\sin\alpha_2 =
0.1$ and $0.5$ respectively}
\label{figmix}
\end{center}
\end{figure}

Fig.~\ref{figmixgghscan} presents the overall dependence on
$\alpha_2$. The strong oscillations might seem surprising.
To explain them on the example of $T_+$, let us first note
that at $\alpha_2\approx\pi/2$ and $\tan\beta\gg 1$ the boson $h_1$
becomes almost pseudoscalar.  Next, it is well known that the
two-photon decay width of the pseudoscalar is significantly smaller
than the $h \to \ggam$ decay width.  Therefore, quantity $T_+$ should
be close to zero at $\alpha_2\approx \pi/2$. For more details, one
can consider this quantity $T_+$ for the case $\tan\beta=3$, for
definiteness. In this case $\sin^2\beta=0.9$. By definition,
$T_+\propto |G|^2 + |\tilde G|^2$, the $W$ contribution in the first
term plays the dominant role everywhere except for narrow region near
$\alpha_2 = \pi/2$ and thus dictates the shape $T_+\propto[1 -
0.9(1-\cos\alpha_2)]^2+$ {\em small remnants}.  At $\alpha_2$
slightly above $\pi/2$, when $t$-quark exactly cancels the remnant of
$W$ boson contribution (and the real part of the $b$ contribution),
$T_+$ is saturated by $|\tilde G|^2$, which is intrinsically smaller
than $|G_{SM}|^2$ by two orders of magnitude. The shape of $T_-$,
etc.  dependence on $\alpha_2$ can also be foreseen from
Eq.(\ref{CPMSM}) in the same way. Our calculations show that the
quantities $T_\perp$, $T_\psi$ as well as asymmetries $V_i$ of the
\egeh\ reaction also exhibit a similar oscillatory dependence on
$\alpha_2$.  The principal features of the results remain the same
for other values of Higgs boson masses, including region
$M_h>2M_W$ above the WW threshold.

However, the case of strong \CP mixing is obviously so prominent that
it will be seen at other colliders.  The opposite case --- the "weak
mixing regime" (small values of $\alpha_2$) --- looks especially
interesting.  The above equations show that in this region
$T_-,\,T_\psi,V_\psi\propto\alpha_2$, $V_\lambda \propto c\lambda_V
+\alpha_2$, all other quantities differ from their values without \CP
mixing only a little, by a quantity $\sim \alpha_2^2$. Therefore, the
asymmetries $T_-,\,T_\psi$ for \ggam\  collisions and
$V_\psi,\,V_\lambda$ for \egam collisions are most sensitive to the
weak \CP mixing, as it is seen in Figures.

The quantities $T_-$ and $T_\psi$ are non-zero only due to \CP
violation. Their $\tan\beta$-dependence for different $\alpha_2$ is
shown in Fig.~\ref{figmix}.  The measurements of both of these
quantities supplement each other essentially: asymmetry $T_\psi$ is
most sensitive to mixing effects at large $\tan\beta$, while in the
small $\beta$ domain the best suited quantity is $T_-$.  This
$\tan\beta$ dependence of the both quantities again can be traced
form Eq.(\ref{CPMSM}). Asymmetry $T_-$, being proportional to
$Re(\tilde G_\gamma G_\gamma^*)$, borrows its $\tan\beta$ behavior
from interplay of the $b$ and $t$ quark contributions to $Re(\tilde
G_\gamma)$: the $b$ contribution, initially small, grows with
$\tan\beta$. It compensates the $t$ loop at $\tan\beta\approx 10$
and becomes dominant later on.  At the same time, $T_\psi$ has
$\tan\beta$--dependence similar to $Im(\tilde G_\gamma)$, where we
have only $b$ quark loop contribution. Thus, the whole asymmetry
$T_\psi$ scales as $\tan\beta$.

For the \egam\ collision we present only the quantities arising
from \CP\ non-conservation, they are $\propto \alpha_2$ at small
$\alpha_2$ (Fig.~\ref{figmix}). Just as for $\ggam$ reaction the
studies of both these quantities supplement each other. The effect
of circular polarization $V_\lambda^L$ (which is an analogue to
$T_-$) is relatively large at $\tan\beta\sim 1$, the $t$/$b$ quark
loop compensation point diminish this effect with growth of
$\tan\beta$ (it becomes zero at large $\tan\beta$).  Thus, in the
whole $\tan\beta$ domain under investigation the $t$ quark loop in
$\tilde G_i$ is dominant and therefore makes $V_\lambda^L$ behave
roughly as $\cot\beta$.  On the contrary, the effect of
linear photon polarization $V_\psi^L$ (which is similar to
$T_\psi$) is very small at $\tan\beta\sim 1$ but it grows with
$\tan\beta$. Nevertheless, it stays below $0.05$ and seems thus
hardly measurable.

The obtained results describe also production of the lightest
Higgs boson in the \MSSM\ in the decoupling regime (when all
superparticles are heavy enough). It is necessary to note in
this respect that the modern calculations in the \MSSM\ need to
fix many subsidiary parameters. In the standard choice, the
variation of Higgs mass and $\tan\beta$ shifts also quantity
$\delta$, so that curves of Ref.~\cite{choi,CDT}, for example,
present simultaneous dependence on parameters $\alpha_2$,
$\beta$ and $\delta$. That is why numerical results of
\cite{choi,CDT} obtained for the specific problems discussed
there differ from our Figs.~\ref{figmixgghscan},\ref{figmix}.
The numerical experiments show that simple variation of \MSSM\
parameters $A$ and $\mu$ allows one to have \SM\ like value
$\sin\delta\approx 1$ at $M_h=105-125$ GeV \cite{gko}. Our
curves correspond to this very case of \MSSM.

\section{Discussion}

In this work, together with \cite{anom1}, we gave detailed answers
to the questions what is the whole experimentally available
information about photon-Higgs boson anomalous interactions and
how to extract it in a reasonable way from future experiments at
Photon Colliders. In this problem, the comparative simultaneous
analysis of both reactions $\ggam \to H$ and \egeh is useful.
Due to the absence of \SM couplings of the Higgs boson with
photons at tree level, the signal of non-standard phenomena can
appear clean in Higgs boson production in photon collisions. The
high sensitivity of reactions $\ggam \to H$ and \egeh\  to the
admixture of various anomalous interactions makes these
processes very useful in exploring the New Physics beyond TeV
scale. With new degrees of freedom (\ref{LEFF}) in the
parametric space, the unique opportunities of Photon Colliders
in the variation of the initial photon polarization provide a
new route to studying different anomalies in details and
confident separation of different contributions.

In our investigation we treat anomalies in a universal manner,
regardless of the particular mechanism of the \CP violation
phenomenon. This is possible because, as we showed, various
sources of \CP violation are indistinguishable in the two
reactions discussed having relatively large cross sections.
These mechanisms are, in principe, distinguishable via the study
of such processes as $\ggam\to HH$ or $\ggam\to H^*\to ZZ$ at
$s\gg M_H^2$. However they have very low cross sections and will
hardly help.

Aiming at the most wide class of anomalous interactions, we
parameterized the amplitudes in a very general way, treating the
absolute values $|g_i|$ and phases $\xi_i$ of anomalies as
independent parameters. The results presented shows the range of
effects that could be resolved from the data, it is close to that
for the \CP-even case \cite{anom1}. They are $g_\gamma,\,
g_{P\gamma} \sim 0.5\div 1\cdot 10^{-4}$ for $H\ggam$ anomalies
and $g_{Z},\, g_{PZ} \sim 5\cdot 10^{-4}$ for $HZ\gamma$ anomalies
(in terms of $\Lambda_i$ introduced in \cite{anom1} they read
$\Lambda_\gamma,\; \Lambda_{P\gamma}\sim 40\div 60$ TeV and
$\Lambda_Z,\; \Lambda_{PZ}\sim 20$ TeV).  Effects depend strongly
on the phase of anomaly. The comparative study of effects with
circularly and linearly polarized photons is necessary to separate
effects of amplitude and phase of anomaly ($|g_i|$ and $\xi_i$).
Future simulations based on final versions of collider and
detector will show the exact discovery limits before actual
experiments.

Next, we analyzed some specific cases of anomalies: the presence
of new particles within \SM (for \CP even anomalies,
\cite{anom1})\fn{ Note that "existence of extra chiral
generations with all fermions heavier than $M_Z$ is strongly
disfavored by the precision electroweak data. However the data
are fitted nicely even by a few extra generations, if one allows
neutral leptons to have masses close to 50 GeV" \cite{okun1}}
and scalar-pseudoscalar mixing in the \DSM.  Their important
feature is definite relation among the anomalous signals in
\ggam and \egam collisions. In particular, the study of both
\ggam\  and \egam\  reactions is essential to test if we deal
with either \CP violating mixing in 2HDM with definite relation
among $H\ggam$ and $HZ\gamma$ anomalies or with some other
mechanism of \CP violation with now unpredicted relation between
these two anomalies. The specific feature of result is that
signals of small mixing ($\sin\alpha_2\sim 0.1$) are seen well
in effects with circular photon polarization at small and large
$\tan\beta$ (but not at intermediate, $\tan\beta\sim 10$),
whereas effects with linear photon polarization can be seen well
at intermediate and large values of $\tan\beta$.

Last, it is useful to note one more advantage of analysis of
polarization asymmetry in the production of Higgs bosons. There
is a possibility in the \DSM\  and \MSSM\  that the heavier
scalar Higgs boson $H$ and its pseudoscalar counterpart $A$ are
almost degenerate within the mass resolution without \CP
violation. In this case the study of polarization asymmetries
{\em in Higgs boson production} like those discussed above can
answer whether \CP is violated or not. Contrary to this, the
study of asymmetries of {\em decay products} cannot distinguish
the true \CP violation from accidental overlapping of $H$ and
$A$ resonance curves.

We are thankful to V. Ilyin, M. Krawczyk, V. Serbo and P. Zerwas
for discussions. IPI is thankful to Prof.~J.~Speth for
hospitality at Forschungszentrum J\"ulich and IFG is thankful to
Prof.~A.~Wagner for hospitality in DESY, where the paper was
finished. This work was supported by grants RFBR 99-02-17211 and
00-15-96691, grant ``Universities of Russia" 015.0201.16 and
grant of Sankt-Petersburg Center of fundamental studies.

 \end{document}